\documentclass[prl,floatfix,amsfonts,twocolumn,superscriptaddress]{revtex4}  
\usepackage{amsmath}
\usepackage{amssymb}
\usepackage{graphicx}
\usepackage{graphics}
\usepackage{epsfig}
\usepackage{bm}
\usepackage{color}

\newcommand{\tg}{\tilde{g}}

\newcommand{\cA}{{\cal A}}

\newcommand{\balpha}{{\bm \alpha}}

\newcommand{\ba}{{\bm a}}

\newcommand{\bx}{{\bm x}}

\newcommand{\bA}{{\bm A}}

\newcommand{\rA}{{\rm A}}

\newcommand{\rtext}[1]{\textcolor{black}{#1}}
\newcommand{\btext}[1]{\textcolor{black}{#1}}

\begin{document}

\title{\rtext{Stable nonlinear modes sustained by gauge fields}
}

\author{Yaroslav V. Kartashov}
\affiliation{Institute of Spectroscopy, Russian Academy of Sciences, Troitsk, Moscow, 108840, Russia
} 
\author{Vladimir V. Konotop}
 \affiliation{Departamento de F\'{i}sica, Faculdade de Ci\^encias, Universidade de Lisboa, Campo Grande, Edif\'icio C8, Lisboa 1749-016, Portugal and Centro de F\'{i}sica Te\'orica e Computacional, Universidade de Lisboa, Campo Grande, Edif\'icio C8, Lisboa 1749-016, Portugal. 
 }

\date{\today}

\begin{abstract}
We reveal the universal effect of gauge fields on the existence, evolution, and stability of solitons in the spinor multidimensional nonlinear Schr\"{o}dinger equation. Focusing   on the two-dimensional case, we show that when gauge field can be split in a pure gauge and a \rtext{non-pure gauge} generating \rtext{effective potential}, the roles of these components in soliton dynamics are different: the \btext{localization characteristics} of emerging states are determined by the curvature, while pure gauge affects the stability of the modes. Respectively the solutions can be exactly represented as the envelopes independent of the pure gauge, modulating stationary carrier-mode states, which are independent of the curvature. Our central finding is that nonzero curvature can lead to the existence of unusual modes, in particular, enabling stable localized self-trapped fundamental and vortex-carrying states in media with constant repulsive interactions without additional external confining potentials and even in the expulsive external traps.
 
\end{abstract}

\maketitle

Gauge fields are ubiquitous in physical theories ranging from classical electromagnetism~\cite{history}, theories of molecular~\cite{Mead}, spinor~\cite{spintron}, and many-body ~\cite{LittRein} systems, to particle physics~\cite{particle} and gravity~\cite{gravity}. Gauge fields can be artificially created in atomic systems~\cite{GaugeReview2011}, a spin-orbit coupled (SOC) Bose-Einstein condensate (BEC)~\cite{Nature} being a celebrated example, and in photonics~\cite{photon}. Since nonlinearity is naturally present in these systems, the effect of the gauge fields on dynamics of nonlinear waves in them has become a subject of growing interest. New types of spatially localized nonlinear waves - solitons - were introduced in SOC BECs. In one-dimensional (1D) settings solitons with various symmetries imposed by a constant~\cite{Kevrekidis}, spatially dependent~\cite{KarKonZez14,KK,KKMS}, {and nonlinear gauge fields~\cite{chiral},} were proposed in pseudospin-1/2 BECs,  as well in SOC-BECs with integer pseudospin~\cite{Adhikari}.  It was established ~\cite{KK,KKMS} that inhomogeneous gauge fields may be described by an integrable extension of the Manakov model~\cite{Manakov}, i.e., by coupled nonlinear Schr\"{o}dinger (NLS) equations with SU(2)-symmetric nonlinearity. 

Gauge fields in 2D nonlinear dynamics may have remarkable stabilizing effect. Already in \cite{Chick} it was noticed that combination of a rotating trap with external parabolic potential, viewed as a gauge potential, sustains stable solitons of 2D NLS equation. A possibility of confinement by a gauge field in a spinor BEC in the quasi-relativistic limit was mentioned in~\cite{Merkl}. Solitons and half-vortices in SOC BECs were reported in the presence of Zeeman lattices~\cite{Lobanov,SakMalom2016}, in free space~\cite{Stabil2D,KTMSMK}, and in dipolar SOC BECs~\cite{dipolar}. Stable 2D solitons were also found in optical media with dispersive coupling mimicking SOC~\cite{SOC-optics}. Thus, it was shown that SOC by itself can stabilize 2D solitons in media with attractive interactions. Moreover, stable self-trapped modes have been found numerically in SOC-BECs with self- and cross-interactions of different signs~\cite{Malomed2017}. At the same time, spatially localized 2D BEC solitons with all repulsive interactions under the action of gauge fields have been studied only in the presence of either external traps~\cite{KaTsU,Santos} or lattices~\cite{Lobanov}. There exist also theoretical predictions of metastable solitons in 3D SO-BECs sustained by 3D SOC~\cite{Stabil3D}, as well as solitons sustained by 2D SOC in the presence of a Zeeman splitting~\cite{KTMSMK}. 

In this Letter we show that pure gauge and non-pure gauge components of the field have profoundly different effects on nonlinear modes. \rtext{A spatially localized solution of SU(2)-symmetric NLS equations allows for the (exact) representation [see (\ref{factor}) below] in a form  of a {\em carrier state} (determined by the pure gauge) which is modulated by the {\em envelope} evolving in an effective potential induced by the non-pure gauge. Thus, the latter affects the very existence of solitons. In contrast, the pure gauge does not have impact on the existence of nonlinear modes. However, by changing the structure of the carrier states, i.e., the internal structure of solitons, it does affect the soliton stability.} Breaking SU(2) symmetry by nonlinearity reveals stabilizing properties \rtext{of the field that is pure gauge in the SU(2) limit. Exploring the 2D case,}  we illustrate confining and stabilizing (or destabilizing) effects of a gauge field on solitons in the repulsive medium without confining linear potentials, and even in the presence of expulsive traps. \rtext{Remarkably, even vortex solitons can be stabilized by gauge fields}.

\paragraph{The model and gauge transformation--} 
The phenomena described below are governed by the two-component NLS equation:
\begin{equation}
\label{GPE}
i\partial_t\Psi=\frac{1}{2}
\left[-i \nabla+\bA (\bx)\right]^2\Psi +V_0(\bx)\Psi+G\left(\Psi^\dag,\Psi\right)\Psi
\end{equation}
for a spinor $\Psi(\bx,t)=(\psi_1,\psi_2)^{\rm T}$ in a $d-$dimensional space $\bx\in \mathbb{R}^d$. In Eq.~(\ref{GPE}), 
$\bA(\bx)=(\rA_1,...,\rA_d)$ is the vector gauge field, whose entries $\rA_{j}(\bx)$ are Hermitian $2\times 2$ matrices, the matrix $G(\Psi^\dag,\Psi)$  describes the nonlinearity, and $V_0(\bx)$ is the external potential. The gauge field determines the curvature~\cite{YangMills} $ 
F_{mn}(\bA)=\partial_m\rA_n-\partial_n\rA_m-i[\rA_m,\rA_n],
$ where $\partial_n\equiv\partial/\partial x_n$. One can define an \rtext{effective "magnetic" field~\cite{GaugeReview2011}} as $B=\varepsilon_{mn}F_{mn}(\bA)$ and $B_l=\varepsilon_{lmn}F_{mn}(\bA)$, in 2D and 3D cases, respectively, where $\varepsilon$ is the Levi-Civita symbol (here the sum over repeated indexes is computed).

In the linear case ($G=0$) a gauge field generating zero curvature (alias, pure gauge) can be gauged out provided that space domain is simply connected. The possibility to eliminate pure gauge, \btext{using judiciously chosen gauge transformation,} persists also in the nonlinear case, provided the nonlinearity is SU(2)-symmetric~\cite{KKMS}. Suppose that a gauge field generates nonzero curvature, i.e., $F_{mn}(\bA)\neq 0$. One can represent $\bA(\bx)= \tilde{\bA}(\bx)+\ba(\bx)$, where $ \tilde{\bA}$ is a pure gauge, and hence $F_{mn}( \tilde{\bA})=0$ for all $m$ and $n$, while $\ba$ is a component generating non-zero curvature (both $\tilde{\bA}(\bx)$  and $\ba(\bx)$ are Hermitian). Consider also 
 an orthonormal basis, $\Phi_{1,2}(\bx)$ ($\Phi_i^\dagger\Phi_j=\delta_{ij}$)  in the spinor space consisting of the kernel of the operator $-i \nabla+ \tilde{\bA}(\bx)$: 
$i \nabla\Phi_j=\tilde{\bA}\Phi_j$. The existence of such a basis is ensured by the zero-curvature, which is also integrability, condition $F_{mn}( \tilde{\bA})=0$, and by the hermiticity of the components of 
$\tilde{\bA}$~\cite{KKMS}. Now one can define the unitary $2\times 2$ matrix solution $\Phi=(\Phi_1,\Phi_{2})$ ($\Phi^\dagger\Phi=\sigma_0$ where $\sigma_0$ is $2\times 2$ identity matrix) \btext{and perform   the gauge transformation}
\begin{eqnarray}
\label{factor}
\rtext{ 
\Psi=\Phi U, \qquad {U}(\bx, t)=(u_1(\bx,t), u_2(\bx,t))^{\rm T}.
}
\end{eqnarray}
The spinor $U$ solves the equation \begin{eqnarray}
\label{GPE_u}
i\partial_t {U}=-\frac{1}{2}\nabla^2{U}+G({U}^\dagger\Phi^\dagger, \Phi {U}){U}-i\balpha\cdot\nabla {U}  
\nonumber \\
-\frac{i}{2}(\nabla\cdot\balpha){U} +\frac{1}{2}\balpha^2{U}+V_0(\bx){U}
\end{eqnarray}
where \rtext{the transformed gauge field $\balpha=\Phi^\dagger\ba\Phi$ originates an  effective potential}. 
The spinor ${U}$ can be viewed as an {\em envelope}, independent of the pure gauge, propagating against stationary {\em "carrier"} states $\Phi_{1,2}(\bx)$, which do not depend on the curvature. \rtext{It follows from Eq.~(\ref{GPE_u}) that in the case of SU(2)-invariant nonlinearity, $G({U}^\dagger\Phi^\dagger, \Phi {U}) =G({U}^\dagger,{U})$, the presence of a pure gauge, $\bA= \tilde{\bA}$ and $\ba=0$, does not affect the existence of nonlinear modes, although a spatial profile of a solution -- if any solution exists -- depends on the gauge field. If, however, $\ba\neq 0$ the equation for the envelope becomes gauge-field dependent, and thus the very existence of the modes can be affected by such a field.}  For non-SU(2)-invariant nonlinearity, a   pure gauge may affect the existence of stationary solutions, because now the nonlinearity becomes $\bx$-dependent.

The solutions $\Phi_{1,2}$ can be found explicitly in a number of cases. For example, if the components of $ \tilde{\bA}$ are of the form $A_{0m}=X_m(x_m)\cA_m$, where $X_m(x_m)$ are real-valued functions of one variable, and  $\cA_m$ are Hermitian matrices simultaneously diagonalizable by a unitary transformation $S$: $S^{-1}\cA_m S=\mbox{diag}(\lambda_{1m},\lambda_{2m})$, then  
\begin{equation}
\label{Phi}
\Phi_1=\frac{S}{\sqrt{2}}\left(\begin{array}{c}
1\\0
\end{array}\right) e^{-i\sum_j\!\lambda_{1j}\xi_j},\,\,\, \Phi_2=\frac{S}{\sqrt{2}}\left(\begin{array}{c}
0\\1
\end{array}\right) e^{-i\sum_j\!\lambda_{2j}\xi_j}
\end{equation} 
where $\xi_j(x_j)=\int X_j(x_j)d x_j$ are the real functions.

If some spatial components of $\bA$, considered alone, constitute a pure gauge, the spatial dimensionality of $\ba$ can be made less than that of $\bA$, and hence less than $d$. For example, if in the 3D space $A_1=A_1(x_1)$ [$\partial_2A_1=\partial_3A_1=0$] one can choose $ \tilde{\bA}=(A_1,0,0)$, so that $\ba=(0,A_2,A_3)$ is a 2D matrix vector \rtext{in the 3D space}. 
Hence, if \rtext{ there exists} a \rtext{stable} nonlinear mode 
sustained by such gauge field, then there \rtext{must  also} exits a stable mode sustained by a gauge field of a lower dimension (see ~\cite{KTMSMK} for examples) \rtext{which is obtained by gauging out the respective spatial components of $\bA$.}

\paragraph{Two-dimensional models--} Turning now to the 2D case, we present a variety of \rtext{examples of} self-sustained stable modes sustained by the  curvature. To this end we consider $ \tilde{\bA}=(A_1(x_1)\sigma_1, A_2(x_2)\sigma_1) $, where $A_{j}(x_j)$ can be arbitrary functions ensuring the existence of the integrals $\xi_j$ (and of the matrix $\Phi$) defined in (\ref{Phi}). Focusing on the simplest (Abelian) case \rtext{$[A_n,A_m]=0$,} we set $\ba=(\zeta_1(x_2)\sigma_1,\zeta_2(x_1)\sigma_1)$
where functions $\zeta_{j}(x_{3-j})$ are so far arbitrary. In this case $\balpha=-(\zeta_1(x_2)\sigma_3,\zeta_2(x_1)\sigma_3)$
and $\nabla\cdot\ba=0$. We consider typical nonlinearity
\begin{eqnarray}
\label{Manakov}
G(\Psi^\dag,\Psi)=\left(
g|\psi_1|^2+\tg |\psi_2|^2, 
g|\psi_2|^2+\tg|\psi_1|^2
\right)^{\rm T}
\end{eqnarray} 
where $g>0$ ($\tg>0$) and $g<0$ ($\tg<0$) correspond to repulsive and attractive intra- (inter-) species interactions.  If $\tg=g$, i.e., $G(\Psi^\dag,\Psi)=g\Psi^\dag\Psi$, Eq.~ (\ref{GPE}) is unitary equivalent to an SU(2)-invariant spinor NLS equation~\cite{KKMS}. Since the latter does not support stable 2D localized states, a pure gauge also cannot lead to their emergence. However, the picture changes drastically for nonzero curvature $B\neq 0$. 

\begin{figure}
\centering
\includegraphics[width=\linewidth]{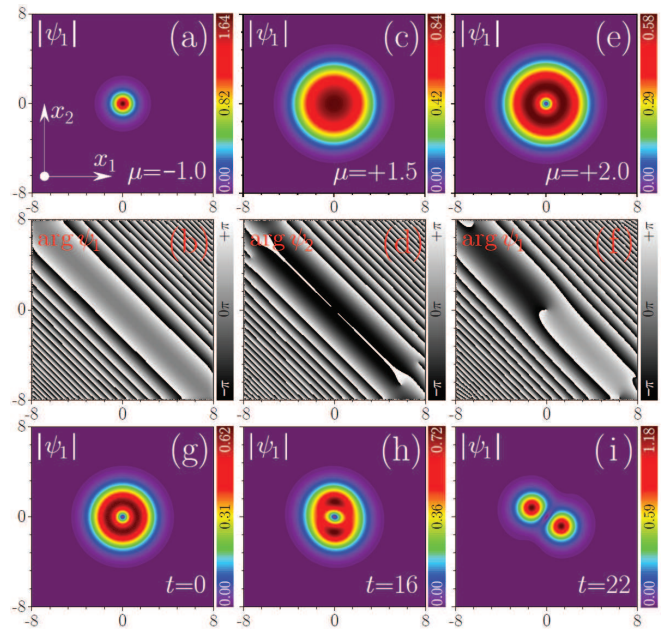}
\caption{Absolute value $|\psi_1|$ (upper row) and phase distributions (middle row) for the fundamental soliton [(a),(b)] in the attractive medium, $g=-1$, and self-trapped fundamental [(c),(d)] and vortex [(e),(f)] pseudo-solitons in the repulsive medium, $g=1$ (corresponding $\mu$ values are indicated in the plots and by red dots in Fig.~\ref{fig:two} below). 
These are stable and have identical $|\psi_1|$ and $|\psi_2|$ distributions,  {but different phases.} (g)-(i) Decay of the unstable vortex soliton with $\mu=1.05$ in the attractive medium.   Numerical results shown here and below were obtained for the original system (\ref{GPE}) and for $B=1$.}
\label{fig:one}
\end{figure}

\begin{figure}
\centering
\includegraphics[width=\linewidth]{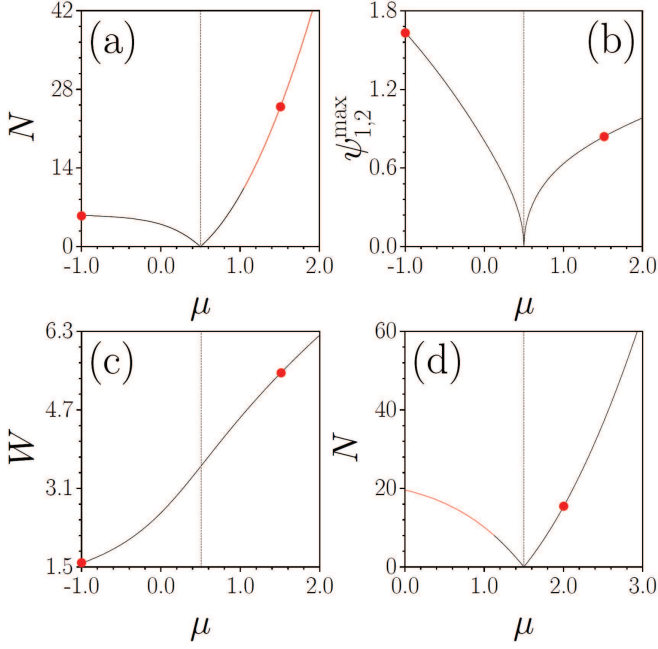}
\caption{Norm $N$ (a), maximal amplitude $\psi^{\rm max}_{1,2}$ (b), and integral width $W$ (c) of the fundamental solitons ($\mu<1/2$, $g=-1$) and pseudo-solitons ($\mu>1/2$, $g=+1$), and norm of vortex solitons and vortex pseudo-solitons (d) {\em vs} $\mu$. The $N(\mu)$ curves for families bifurcating from $\Psi_{1,2}^{\rm lin}$ coincide. The curves in (a) also exactly coincide with those for solitons bifurcating from the linear superposition $\Psi_1^{\rm lin}+\Psi_2^{\rm lin}$. They also coincide with families obtained for non-SU(2)-symmetric nonlinearities $g=-1.4$, $\tg=-0.6$ (at $\mu<1/2$) and $g=1.4$, $\tg=0.6$ (at $\mu>1/2$). The family bifurcating from $\Psi_1^{\rm lin}+\Psi_2^{\rm lin}$, however, is unstable at $\mu\gtrsim 1.05$ [indicated by red curve in (a) coinciding with stable branch for the family bifurcating from $\Psi_1^{\rm lin}$]. Stable (unstable) families are shown black (red).}
\label{fig:two}
\end{figure}

First we consider Eqs. (\ref{GPE}), (\ref{Manakov}) with $\tilde{g}=g$ and without external trapping potential, $V_0(\bx)\equiv 0$. We set $ \tilde{\bA}=\bx$ and explore the curvature introduced by a constant field $B$ generated by $\ba=((1+B/2)x_2\sigma_1,(1-B/2)x_1\sigma_1)$. In Fig.~\ref{fig:one} we show examples of nonlinear states sustained by such curvature \rtext{[all reported numerical results were obtained for (\ref{GPE})]}. A stable fundamental soliton in the medium with attractive nonlinearity is shown in Fig.~\ref{fig:one}(a),(b). More striking results are presented in Fig.~\ref{fig:one}(c)-(f), illustrating stable localized fundamental and vortex-carrying states obtained for repulsive nonlinearity \rtext{(vortices of a single repulsive gauged NLS equation in a constant magnetic field have been reported in~\cite{Barashenkov})}. Obviously, such states would not exist without  curvature at $V_0(\bx)\equiv 0$, so we term them \textit{pseudo-solitons}. Their amplitudes (upper row) feature rotational symmetry, while their phases (middle row) are highly unconventional and have different symmetries for different types of solutions [they are determined by phase structure of $\Phi_1$ or $\Phi_2$ carrier states defined in (\ref{Phi}) as discussed below]. The states in Fig.~\ref{fig:one} exist as stationary families ($\Psi\propto e^{-i\mu t}$) parameterized by the chemical potential $\mu$ determining the norm of the solution $N=\int_{\mathbb{R}^2}\Psi^\dagger\Psi d\bx$. The dependencies $N(\mu)$ for fundamental solitons and pseudo-solitons are presented in Fig.~\ref{fig:two}(a). In the whole domain of $\mu$ shown, the families of fundamental solitons ($\mu<B/2$, attractive medium) and pseudo-solitons ($\mu>B/2$, repulsive medium) are dynamically stable [the red unstable family in panel (a) corresponds to a "superposition" state discussed below]. 

Gauge field can sustain {\em localized linear modes}, i.e., stationary solutions of (\ref{GPE}) at $G\equiv 0$. In our case they are given by $\Psi_{j}^{\rm lin}=  e^{-i(-1)^jx_1x_2-iBt/2 -B\bx^2/4}\Phi_j$  with the carrier modes obtained from (\ref{Phi}) with the proper matrix $S$: $\Phi_j=(1,(-1)^j)^{\rm T}e^{i(-1)^j(x_1+x_2)/2}/\sqrt{2}$  ($j=1,2$).  Both families shown in Fig.~\ref{fig:one}(a), bifurcate from one of the states $\Psi_{j}^{\rm lin}$ (families bifurcating from $\Psi_{1}^{\rm lin}$ and $\Psi_{2}^{\rm lin}$ coincide). Notice that the existence of the linear limit supported by 1D SOC in a 2D BEC requires strong anisotropy of the limiting solution~\cite{KTMSMK}. In our case the soliton amplitude remains radially symmetric at $\Psi_j\to \Psi_{j}^{\rm lin}$, $\mu\to B/2$, reflecting different origin of the phenomenon: now the linear limit is sustained by the curvature induced by the gauge field.

When $\mu\to-\infty$ (attractive medium) the fundamental soliton approaches Townes soliton~\cite{Townes} with the norm $N\to N_\textrm{T}\approx 5.85$ [see Fig.~\ref{fig:two}(a)]. The limit $\mu \to +\infty$ (repulsive medium) is described by the Thomas-Fermi (TF) approximation, which now can be applied even in the absence of external potential. The TF limit for pseudo-solitons belonging to the family bifurcating from $\Psi_1^{\rm lin}$ has the form $\Psi_1^{\rm TF}= e^{ix_1x_2-i\mu t}(\mu/g-B^2\bx^{2}/8g)^{1/2}\Phi_1$ and is valid for $|\bx|<(8\mu)^{1/2}/B$. Maximal amplitude $\psi_{1,2}^{\rm max}=\max_\bx  |\psi_{1,2} |$ of the state in the TF limit is thus $\sim \mu^{1/2}$. This is in qualitative agreement with the increase of soliton amplitude shown in Fig.~\ref{fig:two}(b) for $\mu>1/2$. The tendency for contraction of the nonlinear state in the attractive case and its expansion in the repulsive one is illustrated in Fig.~\ref{fig:two}(c) by the dependence of the integral soliton width $W=N[\int_{\mathbb{R}^2}(\Psi^\dag\Psi)^2d\bx]^{-1/2}$ on $\mu$.

Families of localized {\em vortex} states sustained by the gauge field also have the linear limit. In Fig.~\ref{fig:two}(d) we show a family bifurcating  from the linear state $\Psi_{1}^{\rm vort}=  e^{ix_1x_2-i3Bt/2-B\bx^2/4}(x-iy)\Phi_1$. One of our central results is that all vortex pseudo-solitons (at $\mu>3/2$, $g=-1$) as well as vortex solitons (at $1.125<\mu<3/2$, $g=-1$) were found dynamically stable [black curves in Fig.~\ref{fig:two}(d)]. Vortex solitons with $\mu\lesssim 1.125$ in the attractive medium are unstable [red curve in Fig.~\ref{fig:two}(d)] and usually decay into two rotating fragments that do not fly apart, but perform aperiodic radial oscillations [Fig.~\ref{fig:one}(g)-(i)].

\begin{figure}
\centering
\includegraphics[width=\linewidth]{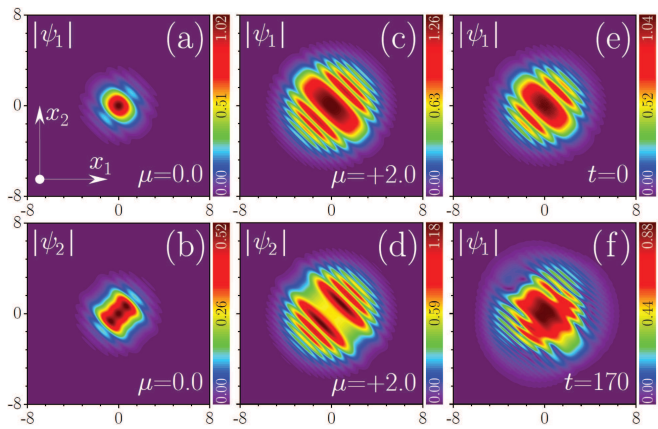}
\caption{Absolute values $|\psi_{1,2}|$ for solitons and pseudo-solitons bifurcating from the  state $\Psi_1^{\rm lin}+\Psi_2^{\rm lin}$ in the attractive (a),(b), $\mu=0$, and repulsive (c),(d), $\mu=2$, media, respectively. (e),(f) Decay of the unstable pseudo-soliton of this type at $\mu=1.4$ in the repulsive medium.}
\label{fig:three}
\end{figure}

\paragraph{A role of a pure gauge.--} Since $ \tilde{\bA}$ can be gauged out and carrier states $\Phi_{1,2}$ are mutually orthogonal, the pure gauge does not affect linear dynamics. Nonlinearity couples the carrier states and since their phases depend on $ \tilde{\bA}$ [see~(\ref{Phi})], it can reveal the presence of pure gauge in two ways. First, nonlinear modes from families bifurcating from a superposition of linear states  $c_1\Psi_1^{\rm lin}+c_2\Psi_2^{\rm lin}$ ($c_{1,2}$ are constants)  display different spatial symmetries and stability properties. In Fig.~\ref{fig:three} we illustrate a stable fundamental soliton   [(a),(b)] and pseudo-soliton  [(c), (d)] belonging to the families  bifurcating from the $\Psi_1^{\rm lin}+\Psi_2^{\rm lin}$ state in the attractive and repulsive media, respectively. The $N(\mu)$ dependence for this family exactly coincides with that shown in Fig.~\ref{fig:one}(a) for solitons bifurcating from either $\Psi_1^{\rm lin}$ or $\Psi_2^{\rm lin}$ states. However, now $|\psi_{1,2}|$ distributions are different: {amplitude distributions in Fig.~\ref{fig:three} reveal the stripe patterns resembling previously known theoretical~\cite{stripe-theor,Santos} and experimental~\cite{stripe-exp} results.} The stability properties are different too: the family of pseudo-solitons becomes unstable at $\mu\gtrsim 1.05$ [red line in Fig.~\ref{fig:two}(a)]. Decay of an unstable pseudo-soliton is illustrated in Fig.~\ref{fig:three}(e),(f).

Second, a gauge field can sustain 2D localized modes also in media with non-SU(2)-symmetric nonlinearity [$\tg\neq g$ in (\ref{Manakov})]. When $\balpha$ is diagonal (as in the case at hand), such a possibility becomes obvious from the transformed equation (\ref{GPE_u}) allowing for "one-component" solutions of either ${U}=(u,0)^{\rm T}$ or ${U}=(0,u)^{\rm T}$ type. For the $\bA$ chosen above, the respective fundamental solitons and pseudo-solitons have amplitude distributions identical to those shown in Fig.~\ref{fig:one}(a),(c) for the same $\mu$ values. The stability of the obtained modes, however is now different. For the particular case of attractive $g=-1,4$, $\tg=-0.6$ and repulsive $g=1,4$, $\tg=0.6$ nonlinearities and for $\ba$ resulting in constant $B$, we found that at $ \tilde{\bA}=\textbf{0}$ all solutions are stable in the range of $\mu$ values shown in Fig.~\ref{fig:one}(a), while for $ \tilde{\bA}=\bx$ pseudo-solitons are unstable in the repulsive medium at $\mu\gtrsim 1.35$.

\begin{figure}
\centering
\includegraphics[width=\linewidth]{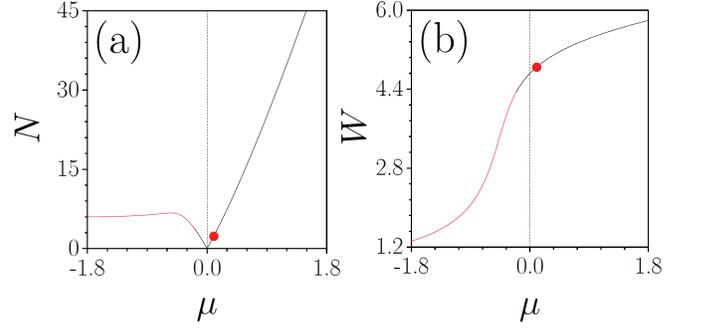}
\caption{ Norm $N$ (a) and integral width $W$ (c) vs $\mu$ for fundamental soliton in the expulsive parabolic trap with $\omega_0=8^{-1/2}$.}
\label{fig:four}
\end{figure}

\begin{figure}
\centering
\includegraphics[width=\linewidth]{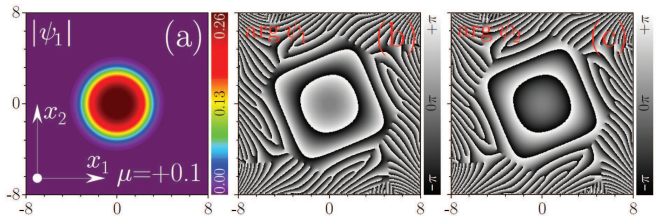}
\caption{A stable pseudo-soliton with $\mu=0.1$ in the repulsive medium with expulsive potential. $|\psi_1|$ and $|\psi_2|$ distributions are identical.}
\label{fig:five}
\end{figure}

\paragraph{Solitons in expulsive trap.--} 
Striking effect of a gauge field on 2D stable localized modes is observed in a medium with either attractive or repulsive SU(2)-invariant nonlinearity in the presence of an {\em expulsive} parabolic trap $V_0(\bx)=-\omega_0^2\bx^2$. We consider spatially-dependent radially symmetric field $B=\omega_0^3\bx^2$ generated by $\ba=(\omega_0^2/3)(-x_2^3,x_1^3)$. Families of the localized solutions, bifurcating from the time-independent linear mode with $\mu=0$ are shown in Fig.~\ref{fig:four}(a). Surprisingly, stable solitons with relatively small amplitudes were found even in the attractive medium, while pseudo-soliton family (repulsive medium) is always stable. Notice that unstable solitons in the attractive medium now may have the norm exceeding norm $N_\textrm{T}$ of the Townes soliton. In Fig.~\ref{fig:five} we show the structure of stable fundamental pseudo-soliton in the expulsive potential. The distributions $|\psi_{1,2}|$ are radially symmetric and identical, while the four-fold symmetry introduced by the gauge field dictates the symmetry of phase distributions of the first [panel 5(b)] and second [panel 5(c)] components. Remarkably, the system supports also solutions with the four-fold symmetric amplitude distributions. For example, a nonlinear family of such solutions bifurcate from a (time-independent) linear state  $\Psi_1^{\rm lin}=(1,-1)^{\rm T}e^{i\bx^2/2-\omega_0^2(x_1^4+x_2^4)/12}$. Vortex solitons and pseudo-solitons exist in expulsive traps as well.

\paragraph{Conclusions.--} We have shown that gauge fields can dramatically affect stability and, when they generate nonzero curvature, the very existence of localized modes. The effect of a gauge field can be "decomposed" into two parts. Pure gauge is responsible for time-independent carrier states and affects soliton stability. The component having nonzero curvature determines the types of the localized solutions and possible evolution scenarios. These phenomena were illustrated on the examples of   two-dimensional solitons in attractive media and pseudo-solitons in repulsive media. Nonlinear modes sustained by gauge fields with nonzero curvature were found even in the presence of repulsive potentials. The approach can be employed for creation of 3D stable localized states,  generating multidimensional modes in systems with integer pseudospin,  investigation of the impact of \rtext{gauge fields on collapse, available so far only for specific 3D models~\cite{DFK}},  etc.


\begin{acknowledgments}
VVK acknowledges financial support from the Portuguese Foundation for Science and Technology (FCT) under Contract no. UIDB/00618/2020. This work was partially supported by the program 1.4 of Presidium of RAS "Topical problems of low temperature physics".
	
\end{acknowledgments}


\begin{thebibliography}{99}
	

\bibitem{history} J. D. Jackson and L. B. Okun, Historical roots of gauge invariance, Rev. Mod. Phys. {\bf 73}, 663 (2001).	

\bibitem{Mead} C. A. Mead, The geometric phase in molecular systems,  Rev. Mod. Phys. {\bf 64}, 51 (1992).
 
\bibitem{spintron} T. Fujita, M. B. A. Jalil, S. G. Tan, and S. Murakami, Gauge fields in spintronics, J. Appl. Phys. {\bf 110}, 121301 (2011). 
 
\bibitem{LittRein} R. G. Littlejohn and M. Reinsch, Gauge fields in the separation of rotations and internal motions in the $n$-body problem, Rev. Mod. Phys. {\bf 69}, 213 (1997).

\bibitem{particle} P. H. Frampton, Gauge Field Theories. (Wiley-VCH, Weinheim, 2008)

\bibitem{gravity} G. Sardanashvily and O. Zakharov, Gauge Gravitation Theory (World Scientific, Singapore, 1992).
 


\bibitem{GaugeReview2011} J. Dalibard, F. Gerbie,  G. Juzeli\~unas, and P. \"Ohberg, Colloquium: Artificial gauge potentials for neutral atoms, Rev. Mod. Phys. {\bf 83}, 1523 (2011);
%
N. Goldman, G. Juzeli\~unas, P. \"Ohberg, and I. B. Spielman,  Light-induced gauge fields for ultracold atoms, Rep. Prog. Phys. {\bf 77}, 126401 (2014).


\bibitem{Nature} Y. J. Lin, K. Jim\'{e}nez-Garc\'{i}a, and I. B. Spielman, Spin-orbit-coupled Bose-Einstein condensates, Nature \textbf{471}, 83 (2011); 
V. Galitski and I. B. Spielman, Spin-orbit coupling in quantum gases, Nature {\bf 494}, 49 (2013). 


\bibitem{photon}
T. Ozawa, H. M. Price, A. Amo, N. Goldman, M. Hafezi, L. Lu, M. C. Rechtsman, D. Schuster, J. Simon, O. Zilberberg, and I. Carusotto, Topological photonics, Rev. Mod. Phys. \textbf{91}, 015006 (2019).

 

\bibitem{Kevrekidis} V. Achilleos, D. J. Frantzeskakis, P. G. Kevrekidis, and D. E. Pelinovsky, Matter-Wave Bright Solitons in Spin-Orbit Coupled Bose-Einstein Condensates, Phys. Rev. Lett. {\bf 110}, 264101 (2013);
Y. Xu, Y. Zhang, and B. Wu, Bright solitons in spin-orbit-coupled Bose-Einstein condensates, Phys. Rev. A {\bf 87}, 013614 (2013).

\bibitem{KarKonZez14}	Y. V. Kartashov,  V. V. Konotop, and D. A. Zezyulin, 
 Bose-Einstein condensates with localized spin-orbit coupling: Soliton complexes and spinor dynamics,  
 Phys. Rev. A {\bf 90}, 063621 (2014). 

\bibitem{KK} 
Y. V. Kartashov and V. V.Konotop, Solitons in Bose-Einstein Condensates with Helicoidal Spin-Orbit Coupling, Phys. Rev. Lett. {\bf 118}, 190401 (2017).
 
\bibitem{KKMS} Y. V. Kartashov, V. V. Konotop, M. Modugno, and E. Ya. Sherman, Solitons in Inhomogeneous Gauge Potentials: Integrable and Nonintegrable Dynamics, Phys. Rev. Lett. {\bf 122}, 064101 (2019).

\bibitem{chiral} {M. J. Edmonds, M. Valiente, G. Juzeli\~unas, L. Santos, and P. \"Ohberg, Simulating an Interacting Gauge Theory with Ultracold Bose Gases, Phys. Rev. Lett. {\bf 110}, 085301 (2013).}

\bibitem{Adhikari} S. Gautam and S. K. Adhikari, Vector solitons in a spin-orbit-coupled spin-2 Bose-Einstein condensate, Phys. Rev. A {\bf 91}, 063617 (2015); Yu-E Li and Ju-K. Xue, Stationary and moving solitons in spin–orbit-coupled spin-1 Bose–Einstein condensates, Front. Phys. {\bf 13}, 130307 (2018);
N.-S. Wan, Yu-E Li, and Ju-K. Xue, Solitons in spin-orbit-coupled spin-2 spinor Bose-Einstein condensates, Phys. Rev. E {\bf 99}, 062220 (2019).

\bibitem{Manakov} S. V. Manakov, 
On the theory of two-dimensional stationary self-focusing electromagnetic waves, Zh. Eksp. Teor. Fiz. {\bf 67}, 543 (1974) [Sov. Phys. JETP {\bf 38}, 248 (1974)].
 

\bibitem{Chick} J. J. Garc\'ia-Ripoll, V. M. P\'erez-Garc\'ia, and V. Vekslerchik, 
Construction of exact solutions by spatial translations
in inhomogeneous nonlinear Schr\"odinger equations, Phys. Rev. A {\bf 64}, 056602 (2001).

\bibitem{Merkl} M. Merkl, A. Jacob, F. E. Zimmer, P. \"Ohberg, and L. Santos, Chiral Confinement in Quasirelativistic Bose-Einstein Condensates, Phys. Rev. Lett. {\bf 104}, 073603 (2010).

\bibitem{Lobanov}  V. E. Lobanov, Y. V. Kartashov, and V. V. Konotop, Fundamental, Multipole, and Half-Vortex Gap Solitons in Spin-Orbit Coupled Bose-Einstein Condensates, Phys. Rev. Lett. {\bf 112}, 180403 (2014).

\bibitem{SakMalom2016} H. Sakaguchi, E. Ya. Sherman, and B. A. Malomed, Vortex solitons in two-dimensional spin-orbit coupled Bose-Einstein
condensates: Effects of the Rashba-Dresselhaus coupling and
the Zeeman splitting, Phys. Rev. E {\bf 94}, 032202 (2016).

\bibitem{KTMSMK} Y. V. Kartashov, L. Torner, M. Modugno, E. Ya. Sherman, B. A. Malomed, and V. V. Konotop, Multidimensional hybrid Bose-Einstein condensates stabilized by lower-dimensional spin-orbit coupling. Phys. Rev. Research {\bf 2}, 013036 (2020).

\bibitem{Stabil2D} H. Sakaguchi, B. Li, and B. A. Malomed, Creation of two-dimensional composite solitons in spin-orbit-coupled self-attractive Bose-Einstein condensates in free space, Phys. Rev. E {\bf 89}, 032920 (2014).

\bibitem{dipolar} {S. Liu, B. Liao, J. Kong, P. Chen, J. L\"u, Y. Li, C. Huang, and Y. Li, Anisotropic Semi Vortices in Spinor Dipolar Bose Einstein Condensates Induced by Mixture of Rashba Dresselhaus Coupling, J. Phys. Soc. Jpn. {\bf 87}, 094005 (2018);
	B. Liao, Y. Ye, J. Zhuang, C. Huang, H. Deng, W. Pang, B. Liu Y. Li, Anisotropic solitary semivortices in dipolar spinor condensates controlled by the two-dimensional anisotropic spin-orbit coupling, Chaos, Solitons and Fractals {\bf 116}, 424 (2018).}

\bibitem{SOC-optics} Y. V. Kartashov, B. A. Malomed, V. V. Konotop, V. E. Lobanov, and L. Torner, Stabilization of spatiotemporal solitons in Kerr media by dispersive coupling, Opt. Lett. {\bf 40}, 1045 (2015).

\bibitem{Malomed2017} Y. Li, Z. Luo, Y. Liu, Z. Chen, C. Huang, S. Fu, H Tan, and B.A. Malomed, Two-dimensional solitons and quantum droplets supported by competing self-and cross-interactions in spin-orbit-coupled condensates, New J. Phys. {\bf 19}, 113043 (2017).

\bibitem{KaTsU} K.  Kasamatsu, M.  Tsubota, and M. Ueda,  Spin textures in rotating two-component Bose-Einstein condensates, Phys. Rev. A {\bf 71}, 043611 (2005);
J. Radi\'c, T. A. Sedrakyan, I. B. Spielman, and V. Galitski, Vortices in spin-orbit-coupled Bose-Einstein condensates, Phys. Rev.  A {\bf 84}, 063604 (2011);
T. Kawakami, T. Mizushima, M. Nitta, and K. Machida, Stable Skyrmions in SU(2) Gauged Bose-Einstein Condensates, Phys. Rev. Lett. {\bf 109}, 015301 (2012).

\bibitem{Santos} 
S. Sinha, R. Nath, and L. Santos, Trapped Two-Dimensional Condensates with Synthetic Spin-Orbit Coupling, Phys. Rev. Lett. {\bf 107}, 270401 (2011).

\bibitem{Stabil3D} Y.-C. Zhang, Z.-W. Zhou, B. A. Malomed, and H. Pu, 
Stable Solitons in Three Dimensional Free Space without the Ground State: Self-Trapped Bose-Einstein Condensates with Spin-Orbit Coupling, 
Phys. Rev. Lett. {\bf 115}, 253902 (2015).

\bibitem{YangMills} C. N. Yang, and R. L. Mills, Conservation of Isotopic Spin and Isotopic Gauge Invariance, Phys. Rev. {\bf 96}, 191 (1954).


\bibitem{Townes} R. Y. Chiao, E. Garmire, and C. H. Townes, Self-Trapping of
Optical Beams, Phys. Rev. Lett. {\bf 13}, 479 (1964).

\bibitem{stripe-theor} {C. Wang, C. Gao, C.-M. Jian, and H. Zhai, H. Spin-orbit coupled spinor Bose–Einstein condensates. Phys. Rev. Lett. {\bf 105}, 160403 (2010); T.-L. Ho and S. Zhang, Bose–Einstein condensates with spin-orbit interaction. Phys. Rev. Lett. {\bf 107}, 150403 (2011); Y. Li, G. I. Martone, L. P. Pitaevskii, and S. Stringari, Superstripes and the Excitation Spectrum of a Spin-Orbit-Coupled Bose-Einstein Condensate. Phys. Rev. Lett. {\bf 110}, 235302 (2013).}

\bibitem{stripe-exp} {J.-r. Li, J. Lee, W. Huang, S. Burchesky, B. Shteynas, F. \c{C}. Top, A. O. Jamison, and W. Ketterle, A stripe phase with supersolid properties in
spin–orbit-coupled Bose–Einstein condensates, Nature {\bf 543}, 91 (2017).}

\bibitem{Barashenkov} \rtext{I. V. Barashenkov and A O Harin, Topological excitations in a condensate of nonrelativistic bosons coupled
to Maxwell and Chern-Simons fields. Phys. Rev. D {\bf 52}, 2471 (1995).}  

\bibitem{DFK} J.-P. Dias, M. Figueira, and V. V. Konotop, Coupled Nonlinear
Schr\"odinger Equations with a Gauge Potential: Existence and
Blowup, Stud. Appl. Math. {\bf 136}, 241 (2015).
 


\end{thebibliography}
\end{document}